\documentclass[12pt,showpacs,showkeys,amsmath,amssymb]{revtex4}
\usepackage{amsmath,amsfonts,amsthm,amscd,amssymb,latexsym}
\usepackage{bm}
\usepackage{dcolumn}
\usepackage{graphicx}
\usepackage{epstopdf}
\usepackage{color}
\usepackage{epsf}
\usepackage{epsfig}
\usepackage{graphicx, epic, eepic, color}

\def\@{\partial_}

\def\negenspace{\kern-1.1em}

\def\sqr#1#2{{\vcenter{\hrule height.#2pt\hbox{\vrule width.#2pt
height#1pt \kern#1pt \vrule width.#2pt}\hrule height.#2pt}}}
\def\square{\mathchoice\sqr64\sqr64\sqr{4.2}3\sqr{3.0}3}

\date{\today}

\begin{document}
\title{Nonlocal Gravity: Damping of Linearized Gravitational Waves}

\author{B. Mashhoon}
\email{mashhoonb@missouri.edu}
\affiliation{Department of Physics and Astronomy,
University of Missouri, Columbia, Missouri 65211, USA}

\begin{abstract} 
In nonlocal general relativity, linearized gravitational waves are \emph{damped} as they propagate from the source to the receiver in the Minkowski vacuum. Nonlocal gravity is a generalization of Einstein's theory of gravitation in which nonlocality is due to the gravitational memory of past events. That nonlocal gravity is dissipative is demonstrated in this paper within certain approximation schemes. The  gravitational  memory drag leads to the decay of the amplitude of gravitational waves given by the exponential damping factor $\exp{(-t / \tau)}$, where $\tau$ depends on the kernel of nonlocal gravity. The damping time $\tau$ is estimated for gravitational waves of current observational interest and is found to be of the order of, or longer than, the age of the universe.
\end{abstract}



\maketitle

\section{Introduction}

A classical nonlocal generalization of Einstein's theory of gravitation~\cite{AE} has been developed in recent papers~\cite{HM1, HM2, HB, BM, CM1, CM2}. In this theory, nonlocality is introduced into gravitation theory via a causal scalar ``constitutive" kernel that acts as the weight function for a certain average of the gravitational field over past events. The gravitational field is represented by a \emph{local} field that satisfies integro-differential field equations. Thus classical nonlocal gravity is Einsteinian general relativity endowed with a certain simple \emph{memory} of past events; this \emph{nonlocal aspect of gravity simulates dark matter}. That is, in this theory, nonlocality is the main source of phenomena associated with what appears as dark matter in astrophysics. To account for the ``flat" rotation curves of spiral galaxies~\cite{RF, RW, SR}, the characteristic length scale of nonlocality is assumed to be of order  $\lambda_0$~=~10 kpc. This is also the scale over which memory of distant events fades. 

It follows from the weak field approximation of general relativity that gravitational waves satisfy the ordinary wave equation and thus propagate in vacuum with the speed of light. However, in the treatment of linearized gravitational waves in nonlocal general relativity, one encounters the following nonlocal wave equation for the free propagation of linearized gravitational radiation~\cite{CM2}
\begin{equation}\label{1}
\square\, h_{ij}+s_{ij}(x)+\int R(x-y)s_{ij}(y)d^4y=0\,,
\end{equation}
where $h_{ij}$ is the amplitude of gravitational radiation and
\begin{equation}\label{2}
 s_{ij}(x)=\frac{1}{c} \int K(|\mathbf{x}-\mathbf{y}|, \mathbf{x}-\mathbf{y})\frac{\partial h_{ij}}{\partial t}(ct-|\mathbf{x}-\mathbf{y}|, \mathbf{y})d^3y\,.          
\end{equation} 
Here  $x$ represents an event in spacetime with coordinates $x^\mu=(ct, \mathbf{x})$ and $\square :=\eta^{\alpha \beta}\partial_\alpha \partial_\beta$ is the d'Alembertian (wave) operator.  Greek indices run from 0 to 3, while Latin indices run from 1 to 3. Moreover, $g_{\mu \nu}=\eta_{\mu \nu}+h_{\mu \nu}$ is the metric tensor of spacetime and the Minkowski metric tensor  $\eta_{\alpha \beta}$ is given by diag$(-1,1,1,1)$ in our convention. We use units such that $c=1$, unless otherwise specified. The freedom in the choice of coordinates, which translates into the gauge freedom of dimensionless gravitational potentials $h_{\mu \nu}$, can be employed to set $h_{0\mu}=0$, impose the transverse gauge condition $\partial h^{ij}/ \partial x^j=0$ and render $h_{ij}$ traceless~\cite{CM2}. The gravitational wave amplitude $h_{ij}$ thus characterizes the deviation of the spatial metric away from flat 3D Euclidean space. 

It should be emphasized that our way of introducing nonlocality into general relativity is by no means unique; however, it is the simplest approach to nonlocal general relativity that is widely consistent with observation~\cite{HM1, HM2, HB, BM, CM1, CM2}. 

In the general linear approximation of nonlocal gravity, the scalar constitutive kernel of the theory is a \emph{universal} function given by the causal convolution kernel $K(x-y)$. Its reciprocal $R(x-y)$ is a causal convolution kernel as well, as discussed in detail in Ref.~\cite{CM2}. The reciprocity between $K$ and $R$ implies that 
\begin{equation}\label{3}
 K(x-y)+R(x-y)+ \int K(x-z)R(z-y)d^4z=0\,.
\end{equation}
It is interesting to note that $K$ and $R$ can be interchanged in the integrand of Eq.~\eqref{3} by simply changing the variable $z$ to $z'$ given by $z'=x+y-z$. We find it useful to work in the Fourier domain, which is permissible for the functions under consideration here~\cite{CM1}. Thus let
\begin{equation}\label{4}
\hat{f} (\xi) =  \int f(x) e^{-i \xi \cdot x}~ d^4x\,
\end{equation}
be the Fourier transform of $f$, where $\xi \cdot x := \eta_{\alpha \beta}\xi^\alpha x^\beta$. Then, 
\begin{equation}\label{5}
f(x) = \frac{1}{(2\pi)^4} \int \hat{f} (\xi) e^{i \xi \cdot x}~ d^4\xi\,.
\end{equation}
It follows from Eq.~\eqref{3} and the convolution theorem that in the 4D Fourier domain, 
\begin{equation}\label{6}
\hat{K}+ \hat{R} + \hat{K} \hat{R}=0\,.
\end{equation}
Given $K$, one can therefore determine $R$ using Fourier transforms and vice versa. These kernels have dimensions of (length)$^{-4}$, so that $\hat{K}$ and $\hat{R}$ are dimensionless. 

For the reciprocal kernel $R$, we adopt the expression derived in Ref.~\cite{CM2}, namely, 
\begin{equation}\label{7}
R(x-y)=H(x^0-y^0-|\mathbf{x}-\mathbf{y}|)A~e^{-A(x^0-y^0)} q(\mathbf{x}-\mathbf{y})\,.
\end{equation}
Here $H(s)$ is the Heaviside unit step function such that $H(s)=1$ for $s\ge0$ and $H(s)=0$ for $s<0$,  $1/A$ is a constant length such that $0<A \lambda_0 < 1$, where $\lambda_0$~=~10 kpc is the basic nonlocality scale in this theory, and $q>0$ is the reciprocal Newtonian kernel. This reciprocal kernel occurs in the Poisson equation for the gravitational potential $V$ of nonlocal gravity in the Newtonian regime, namely, 
\begin{equation}\label{7a}
  \nabla^2V (\mathbf{x}) = 4\pi G\Big[\rho(\mathbf{x})+  \int q(\mathbf{x}-\mathbf{y}) \rho(\mathbf{y})d^3y\Big]\,.
\end{equation}
Thus $V$ is due to matter of density $\rho$ and ``dark matter" of density $\rho_D$, which is the convolution of $\rho$ with the reciprocal Newtonian kernel $q$; that is, nonlocality simulates dark matter~\cite{RF, RW, SR}. The reciprocal kernel $q$ has been discussed extensively in Ref.~\cite{CM1}, where two possible examples  were worked out in detail, namely,
\begin{equation}\label{7b}
q'(\mathbf{r})=\frac{1}{4\pi \lambda_0}~ \frac{1+A(a+r)}{(a+r)^2}~e^{-A r}\,,
\end{equation}
\begin{equation}\label{7c}
q''(\mathbf{r})=\frac{1}{4\pi \lambda_0}~ \frac{1+A(a+r)}{r(a+r)}~e^{-A r}\,.
\end{equation}
Here $r=|\mathbf{r}|$ and $a$,  $a/ \lambda_0 \ll 1$, is a constant length.  The kernels $q'$ and $q''$ are real positive functions that are integrable as well as square integrable over all space; moreover,  we will assume here that their Fourier transforms are real \emph{positive} functions as well. This was proved for $\hat{q}''$ in Ref.~\cite{CM1}, but the situation regarding $\hat{q}'$ is more complicated. Further discussion of this issue is contained in section V, where it is argued that $\hat{q}'$ is positive for sufficiently small values of $a/\lambda_0$. In fact, it follows from the numerical results presented in figure 1 of Ref.~\cite{CM1} that $\hat{q}' > 0$ for $a/\lambda_0=0.001$. Therefore, we will assume in this paper that $a/\lambda_0$ is always so small compared to unity that $\hat{q}'$ is positive.   

The values assigned in Ref.~\cite{CM1} to the constant parameters $\lambda_0$, $A$ and $a$ of kernel~\eqref{7}, namely, $\lambda_0$~=~10 kpc, $A\lambda_0=0.1$ and $a/\lambda_0=0.001$, are rather tentative and will be used in this paper for some of the numerical estimates in section V. In general, $\lambda_0$ should be a galactic length scale to account for the rotation curves of spiral galaxies, while $A\lambda_0$ and $a/\lambda_0$ should be positive and small compared to unity such that $0<aA\ll 1$. 

Let us note here that $R$ is a positive function that is integrable as well as square integrable and satisfies the causality requirement, namely, it is nonzero only when $x^\mu-y^\mu$ is a future directed timelike or null vector in Minkowski spacetime. This means that  event $y$ must be within or on the past light cone of event $x$ such that $x^0\ge y^0+|\mathbf{x}-\mathbf{y}|$, where $|\mathbf{x}-\mathbf{y}|$ can be characterized as \emph{retardation}.  The calculation of $\hat{R}$ becomes much simplified if we neglect retardation; then, 
\begin{equation}\label{8}
\hat{R}(\omega, \mathbf{k}) \approx \frac{A}{A-i\omega}~\hat{q} (\mathbf{k})\,,
\end{equation}
where we henceforth use the approximation sign instead of the equality sign to indicate that the calculation neglects retardation and is therefore not exact. Ref.~\cite{CM2} should be consulted for a detailed discussion of the approximation scheme that is based on neglecting retardation. Using Eq.~\eqref{6} and inverse Fourier transformation, we find that~\cite{CM2}
\begin{equation}\label{9}
K(x^0, \mathbf{x}) \approx -\frac{A}{(2\pi)^3}H(x^0) \int \hat{q} (\mathbf{k}) e^{i \mathbf{k} \cdot \mathbf{x}}e^{-A(1+\hat{q})x^0}~ d^3k\,.
\end{equation}

The choice of a particular gauge for the gravitational potentials together with the requirement of causality implies that the explicit form of our nonlocal gravitational wave equation is valid in a particular inertial frame and violates time-reversal invariance. 

To understand the physical import of Eqs.~\eqref{1} and~\eqref{2}, a mechanical analogy turns out to be quite useful: These equations are reminiscent of the equation of motion of a linear oscillator with a dissipation term that is proportional to the velocity of the oscillator. Here $\partial h_{ij}/\partial t$ is suggestive of the ``velocity" of the oscillator. With the appropriate sign for the coefficient of the dissipation term, one has a \emph{damped oscillator}. In a similar way, with the proper functional forms for the nonlocal kernels,  Eqs.~\eqref{1} and~\eqref{2} indicate free propagation of gravitational waves with \emph{damping}. The nonlocality of the theory originates from a certain average over past events in spacetime; this memory of the past thus appears to act as a drag that dampens the free propagation of linearized gravitational waves. In nonlocal gravity, memory fades exponentially for events that are distant in space and time; see Eqs.~\eqref{7}, \eqref{7b} and~\eqref{7c}. Similarly, the amplitude of gravitational radiation decays exponentially with time as $\exp{(-t/\tau)}$, where the damping time $\tau$ is related to the nonlocal kernel. 

In general, as the waves propagate freely through Minkowski vacuum, the wave amplitude may grow or decay in time due to nonlocality. We expect that, with the correct nonlocal kernel, the solutions of our linear nonlocal homogeneous wave equation are well behaved and decay in time  leading to the stability of Minkowski spacetime under small perturbations. We show in sections II and III that there is no instability and the waves indeed decay in time \emph{when retardation is neglected in the kernel}; that is, employing the same approximation scheme as in Ref.~\cite{CM2} and using the proper physically reasonable nonlocal kernel, we find that  the waves are exponentially damped. A simplified 2D toy model is discussed in section IV, where exponential damping is demonstrated even in the presence of retardation. It thus appears from the physical arguments provided in this paper that damping of gravitational waves is a feature of nonlocal gravity; however, the general mathematical problem involving retardation remains unsolved. The damping time $\tau$ is estimated in section V for the case of gravitational waves that are of current observational interest. Section VI contains a brief discussion of our results.

\section{Nonlocal Wave Equation}

Let us now concentrate on a component of the gravitational wave amplitude and replace Eqs.~\eqref{1} and~\eqref{2} by the equivalent set
\begin{equation}\label{10}
\square\, \psi+{\cal S}(x)+\int R(x-y){\cal S}(y)d^4y=0\,
\end{equation}
and
\begin{equation}\label{11}
 {\cal S}(x)=\frac{\partial}{\partial t}\int K(|\mathbf{x}-\mathbf{y}|, \mathbf{x}-\mathbf{y})~\psi(t-|\mathbf{x}-\mathbf{y}|, \mathbf{y})~d^3y\,.          
\end{equation} 
As these equations are linear in $\psi$ and the kernels are real, it is convenient to work with a complex wave amplitude $\psi$ with the understanding that the real part of $\psi$ is physically significant. We then look for a solution of the form
\begin{equation}\label{12}
\psi (x)=e^{-i \alpha t} \phi(\mathbf{x})\,,
\end{equation}
where $\alpha$ is in general complex. Substituting our ansatz in Eqs.~\eqref{10} and~\eqref{11}, we find 
\begin{equation}\label{13}
 (\nabla^2+\alpha^2)\phi (\mathbf{x}) +i \alpha \int W_{\alpha}(\mathbf{x}-\mathbf{y})\phi (\mathbf{y})~d^3y=0\,,          
\end{equation} 
where $W_{\alpha}$ is given by
\begin{equation}\label{14}
W_{\alpha}(\mathbf{z}):= F_{\alpha} (\mathbf{z}) +\frac{A}{A-i\alpha}\int e^{(i\alpha -A)|\mathbf{z}-\mathbf{u}|}q(\mathbf{z}-\mathbf{u})F_{\alpha} (\mathbf{u})d^3u\,
\end{equation}
and
\begin{equation}\label{15}
 F_{\alpha} (\mathbf{z}) :=-e^{i \alpha |\mathbf{z}|}K(|\mathbf{z}|, \mathbf{z})\,.          
\end{equation} 
The main issue here is whether all solutions of Eqs.~\eqref{13}--\eqref{15} that satisfy proper boundary conditions are such that $\alpha$, 
\begin{equation}\label{16}
\alpha=\omega + i \Delta\,,
\end{equation}
has a negative imaginary part $\Delta <0$. In this case, 
\begin{equation}\label{17}
\psi (x)=e^{-i \omega t} \phi(\mathbf{x})~e^{\Delta t}\,
\end{equation}
will exponentially \emph{decay} in time. Otherwise, the perturbation will blow up as $t \to \infty$, which is physically unacceptable, as it would indicate an intrinsic instability of Minkowski spacetime within the framework of nonlocal gravity. 

In the Fourier domain, if $\hat{\phi}(\mathbf{k})$ is nonzero, Eq.~\eqref{13} can be written as 
\begin{equation}\label{18}
\alpha^2- |\mathbf{k}|^2 +i\alpha \hat{W_{\alpha}}(\mathbf{k})=0\,,
\end{equation}
which expresses the dispersion of gravitational waves due to nonlocality. Writing $\alpha$ as in Eq.~\eqref{16} and introducing the real $(2{\cal R})$ and imaginary $(2{\cal I})$ parts of $ \hat{W_{\alpha}}$,
\begin{equation}\label{19}
 \hat{W_{\alpha}}=2({\cal R} +i{\cal I})\,,
\end{equation}
we find that Eq.~\eqref{18} reduces to its real and imaginary components
\begin{equation}\label{20}
\omega^2-\Delta^2- |\mathbf{k}|^2 -2(\omega {\cal I} +\Delta {\cal R})=0\,
\end{equation}
and 
\begin{equation}\label{21}
\omega \Delta +(\omega {\cal R}-\Delta {\cal I})=0\,.
\end{equation}
It follows from Eq.~\eqref{21} that 
\begin{equation}\label{22}
\Delta = - \frac{\omega {\cal R}}{\omega - {\cal I}}\,,
\end{equation}
which can be substituted in Eq.~\eqref{20}. In the resulting equation, we can choose $\omega - {\cal I}$ as a new variable and after some algebra we find 
\begin{equation}\label{23}
\omega = {\cal I}\pm \Big ({\cal J} +\sqrt{{\cal J}^2+{\cal R}^2{\cal I}^2 }\Big )^{1/2}\,,
\end{equation}
where ${\cal J}$ is given by 
\begin{equation}\label{24}
{\cal J}= \frac{1}{2}\Big ( |\mathbf{k}|^2-{\cal R}^2+{\cal I}^2\Big )\,.
\end{equation}
If $\Delta$ turns out to be negative independently of the sign of $\omega$, then the solution of the nonlocal wave equation will decay in time and could be physically acceptable. As discussed in the following section,  thus far it has been possible to show this for Eqs.~\eqref{13}--\eqref{15} only in the approximation that retardation is neglected.

\section{Damping}

Consider the approximation scheme, introduced in Ref.~\cite{CM2}, that involves neglecting the retardation in Eqs.~\eqref{14} and~\eqref{15}. This means that  Eq.~\eqref{15} is replaced by $F_{\alpha} (\mathbf{z}) \approx -K(0, \mathbf{z})$. Taking Eq.~\eqref{9} into account, we find that in this approximation $F_{\alpha} (\mathbf{z})$ is independent of $\alpha$, since 
\begin{equation}\label{25}
 F_{\alpha} (\mathbf{z}) \approx A q(\mathbf{z})\,.          
\end{equation} 
Similarly, Eq.~\eqref{14} is approximated by 
\begin{equation}\label{26}
W_{\alpha}(\mathbf{z}) \approx A q(\mathbf{z}) +\frac{A^2}{A-i\alpha}\int q(\mathbf{z}-\mathbf{u})q(\mathbf{u})d^3u\,,
\end{equation}
so that its Fourier transform can be expressed as 
\begin{equation}\label{27}
 \hat{W_{\alpha}}(\mathbf{k}) \approx  A \hat{q}(\mathbf{k})\Big[1+\frac{A \hat{q}(\mathbf{k})}{A-i\alpha}\Big]\,.
\end{equation}
We recall that $A$ is a positive parameter, $0<A\lambda_0 < 1$, and $\hat{q}(\mathbf{k})>0$ by assumption. It is shown in section V that $\hat{q}(\mathbf{k})$ is a function of $|\mathbf{k}|$ and is real and \emph{positive} when $a/\lambda_0$ is sufficiently small compared to unity. It follows from Eq.~\eqref{19} that
\begin{equation}\label{28}
 {\cal R} \approx \frac{1}{2}~A \hat{q}~\Big[1+\frac{A(A+\Delta)}{(A+\Delta)^2+\omega^2}~\hat{q}\Big]\,
\end{equation}
and 
\begin{equation}\label{29}
{\cal I} \approx  \frac{1}{2}~\frac{\omega}{(A+\Delta)^2+\omega^2}~(A \hat{q})^2\,.
\end{equation}
Substituting these results in Eq.~\eqref{22}, we find after some algebra that $\Delta$ can be expressed as 
\begin{equation}\label{30}
 \Delta \approx  -\frac{1}{2}~A \hat{q}~\Big[1+\frac{A^2}{(A+\Delta)^2+\omega^2}~\hat{q}\Big]\,,
\end{equation}
which means that $\Delta <0$ and all the modes decay regardless of the value of $\omega$. Indeed, Eq.~\eqref{30} can be expressed as a cubic equation for $\Delta$ with all positive coefficients, since $A>0$ and $\hat{q}>0$; therefore, there is at least one real root, which must be negative, as it follows from Descartes' rule of signs that the cubic equation cannot have any positive root for $\Delta$. 

We remark here for the sake of completeness that the substitution of Eqs.~\eqref{28} and~\eqref{29} in Eq.~\eqref{23} results in a second  formula involving $\omega$ and $\Delta$. This equation and Eq.~\eqref{30} then constitute two coupled algebraic equations for the two unknowns $\omega$ and $\Delta$; in principle, one can determine  $\omega$ and $\Delta$ in this way in terms of $|\mathbf{k}|$, $\hat{q}(\mathbf{k})$ and $A$. 

To what extent does the result  that $\Delta$ is negative depend upon neglecting retardation? It is shown within the framework of a simple 2D toy model in the next section that all the modes decay even in the presence of retardation.

\section{Toy Model}

Consider the 2D nonlocal wave equation given by
\begin{equation}\label{31}
\square\,\Psi-\epsilon \frac{\partial}{\partial t}\int \chi(x-y) \Psi(t-|x-y|, y)dy=0\,,
\end{equation}
where $\epsilon$, $0<\epsilon \ll1$, is a small parameter. Working to first order in $\epsilon$, we wish to show that all the modes decay provided kernel $\chi(z)$ is the Fourier transform of an even positive function $\hat{\chi}(\zeta)$, 
\begin{equation}\label{32}
\hat{\chi}(\zeta)= \hat{\chi}(-\zeta) > 0\,,
\end{equation}
which implies, among other things, that the function $\chi(z)$ is real and even. Therefore, 
\begin{equation}\label{32a}
\hat{\chi}(\zeta)=  2\int_0^{\infty}\chi(z)\cos{(\zeta z)}~dz\,, \quad \chi (z) =\frac{1}{\pi}\int_0^{\infty} \hat{\chi}(\zeta)\cos{(\zeta z)}~d\zeta\,.
\end{equation}
The assumptions regarding kernel $\chi$ are reminiscent of the fact that in nonlocal gravity, kernel $q$, for instance, is such that $\hat{q}>0$ is just a function of $|\mathbf{k}|$; in fact, kernel $\chi$ is the 2D analog of $Aq$ in Eq.~\eqref{25} of the previous section. 

We assume, as before, that 
\begin{equation}\label{33}
\Psi (t, x)=e^{-i \alpha t} \Phi(x)\,,
\end{equation}
so that Eq.~\eqref{31} takes the form
\begin{equation}\label{34}
\frac{d^2 \Phi(x)}{dx^2}+\alpha^2\Phi (x) +i \epsilon \alpha \int I(x-y)\Phi(y)~dy=0\,,          
\end{equation} 
where 
\begin{equation}\label{35}
I(z) :=e^{i \alpha |z|} \chi (z)\,.          
\end{equation} 
Nonlocal wave equations of the general type of Eq.~\eqref{34} have been the subject of previous investigations; see, for instance, Ref.~\cite{BA} and the references cited therein. 

Let $\hat{\Phi}(k)$ be the Fourier transform of $\Phi(x)$; then, it follows from Eq.~\eqref{34} that if $\hat{\Phi}(k)$ is nonzero, we have 
\begin{equation}\label{36}
\alpha^2- k^2 +i\epsilon \alpha \hat{I}(\alpha, k)=0\,,
\end{equation}
where
\begin{equation}\label{37}
 \hat{I}(\alpha, k)=\int \chi(z)e^{i\alpha |z|-ikz}~dz\,.
\end{equation}
Working to first order in $\epsilon$ in the Fourier domain, it is clear from the treatment of section II that for $\alpha=\omega+i\Delta$, we have
\begin{equation}\label{38}
\omega= \pm k+\frac{1}{2}\epsilon \hat{I}_I(\alpha_0, k)\,, \quad  \Delta=-\frac{1}{2}\epsilon\hat{I}_R(\alpha_0, k)\,,
\end{equation}
where $\hat{I}_I$ and $\hat{I}_R$ are the imaginary and real parts of $\hat{I}$, respectively, and $\alpha_0 :=\pm k$ is the solution of Eq.~\eqref{36} for $\epsilon=0$. To prove that to first order in $\epsilon$ all the modes decay, we need to show that $\hat{I}_R(\alpha_0, k)>0$. 

To this end, we first note that the real part of Eq.~\eqref{37} for $\alpha=\alpha_0$ involves $\cos{(\alpha_0 |z| -kz)}$, since $\chi(z)$ is real; hence, 
\begin{equation}\label{39}
 \hat{I}_R(\alpha_0, k)=\int_{-\infty}^{0} \chi(z)\cos{(\alpha_0+k)z}~dz+\int_{0}^{\infty} \chi(z)\cos{(\alpha_0-k)z}~dz\,.
\end{equation}
Changing the variable $z$ to $-z$ in the first integral, we find
\begin{equation}\label{40}
 \hat{I}_R(\alpha_0, k)=\int_{0}^{\infty} \chi(z)\Big[\cos{(\alpha_0+k)z}+\cos{(\alpha_0-k)z}\Big]~dz\,.
\end{equation}
Substituting either $k$ or $-k$ for $\alpha_0$ in Eq.~\eqref{40}, the integrand remains the same; hence, 
\begin{equation}\label{41}
 \hat{I}_R(\alpha_0, k)=\int_{0}^{\infty} (1+\cos{2kz})\chi(z)~dz\,.
\end{equation}
It follows from this result and Eq.~\eqref{32a} that 
\begin{equation}\label{42}
 \hat{I}_R(\alpha_0, k)=\frac{1}{2}\Big[\hat{\chi}(0)+\hat{\chi}(2k)\Big]\,.
\end{equation}
But $\hat{\chi}$ is positive by assumption, hence 
\begin{equation}\label{43}
 \Delta=-\frac{1}{4}\epsilon \Big[\hat{\chi}(0)+\hat{\chi}(2k)\Big] <0\,,
\end{equation}
which means that to first order in $\epsilon$ all the modes indeed decay even in the presence of retardation. 

We now return to Eq.~\eqref{30} and use this result to estimate the nonlocality-induced damping time for gravitational radiation. 

\section{Damping Time $\tau$}

It follows from our approximate treatment in Eq.~\eqref{30} that the damping time $\tau :=-1/ \Delta$ depends upon $\hat{q}(\mathbf{k})$ as well as the frequency of radiation $\omega$. It is intuitively clear that in nonlocal gravity, the propagation of gravitational waves with wavelengths comparable to, or longer than, the basic length scale $\lambda_0 = 10$ kpc could be significantly affected by nonlocality. In particular, a rough estimate for $\tau$ involving gravitational radiation of wavelength comparable to the nonlocality length scale $\lambda_0$ would be a damping time of order $\lambda_0/c \approx 3 \times 10^{4}$ yr. On the other hand, various observational efforts are under way to detect gravitational waves with wavelengths that are much shorter than $\lambda_0$~\cite{Ri}. It turns out that in this case $\omega \approx \pm |\mathbf{k}|$ and that nonlocality generates only a small perturbation on the propagation of such waves~\cite{CM2}.  In particular, current observational possibilities involve gravitational waves in the frequency range $\nu \gtrsim 10^{-8}$ Hz, where $2\pi \nu \approx |\mathbf{k}|$; therefore, $ \lambda_0|\mathbf{k}| \gtrsim 6\times 10^{4}$ for wave vectors of current experimental interest~\cite{Ri}. For such radiation, $A/|\mathbf{k}| \lesssim 10^{-5}$, since $A\lambda_0 < 1$, and to calculate $\tau$ from Eq.~\eqref{30} we need a proper estimate for $\hat{q}(\mathbf{k})$. In fact, for the cases of observational interest, $\hat{q}(\mathbf{k})$ is small compared to unity. Moreover, the quantity in square brackets in Eq.~\eqref{30} is less that $1+(A/\omega)^2\hat{q}$, which is thus very nearly equal to unity. Therefore, it follows from Eq.~\eqref{30} that 
\begin{equation}\label{43a}
\tau \approx \frac{2}{A\hat{q}(\mathbf{k})}\,
\end{equation}
for the frequency range that is the focus of observational searches for gravitational waves at the present time. 

The Newtonian regime of nonlocal gravity is similar to the phenomenological Tohline-Kuhn approach to modified gravity~\cite{T, K, B}; in particular, the reciprocal kernel $q(\mathbf{r})$ is a generalization of the Kuhn kernel $Q(\mathbf{r})$, 
\begin{equation}\label{44}
Q(\mathbf{r})=\frac{1}{4\pi\lambda_0}~\frac{1}{r^2}\,.
\end{equation}
In fact, two examples of $q(\mathbf{r})$, namely, $q'$ and $q''$ given respectively by Eqs.~\eqref{7b} and~\eqref{7c}, have been constructed starting from $Q(\mathbf{r})$ in Ref.~\cite{CM1} by introducing positive constants $a$ and $A$ to change the functional form of the Kuhn kernel for $r \to 0$ and $r \to \infty$, respectively. It has been shown in Ref.~\cite{CM1} via general arguments that the Fourier transforms of $q'$ and $q''$ are actually dimensionless functions of $|\mathbf{k}|$ and have the following properties: $\hat{q}'' > 0$, $\hat{q}'' > \hat{q}'$ and $\hat{q}' > -a/\lambda_0$; moreover, $\hat{q}'$ and $\hat{q}''$ both vanish as $|\mathbf{k}| \to \infty$. On the other hand, from the expressions for $q'$ and $q''$ given respectively by Eqs.~\eqref{7b} and~\eqref{7c}, it is possible to compute explicitly the Fourier transforms of these functions using  
\begin{equation}\label{45}
\hat{q}(\mathbf{k})=\frac{4\pi}{|\mathbf{k}|}\int_0^{\infty}rq(\mathbf{r})\sin(|\mathbf{k}| r)dr\,. 
\end{equation}

Let us first  consider the special limiting case of $a=0$; then, $q'$ and $q''$ coincide and the resulting reciprocal Newtonian kernel $q_0$,
\begin{equation}\label{45a}
q_0(\mathbf{r})=\frac{1}{4\pi \lambda_0}\Big( \frac{1}{r^2}+\frac{A}{r}\Big)~e^{-A r}\,,
\end{equation}
has the same short distance behavior as the Kuhn kernel. It is straightforward to compute the Fourier transforms of $Q$ and $q_0$ using Eq.~\eqref{45} and the results are
\begin{equation}\label{45b}
\hat{Q}=\frac {\pi}{2\lambda_0|\mathbf{k}|}\, 
\end{equation}
 and
\begin{equation}\label{46}
\hat{q}_0=\frac{A}{\lambda_0(A^2+|\mathbf{k}|^2)}+\frac{1}{\lambda_0|\mathbf{k}|} \arctan{\Big(\frac{|\mathbf{k}|}{A}\Big)}\,. 
\end{equation}
Here we have used the fact that for real values of the constants $p_1$ and $p_2$,
\begin{equation}\label{46a}
\int_0^{\infty}e^{-p_{1}x}\sin{(p_2x)}\frac{dx}{x}=\arctan{(\frac{p_2}{p_1})}\,, \quad p_1\ge 0\,; 
\end{equation}
see formulas 3.941 on page 489 of Ref.~\cite{G+R}.
Thus $\hat{q}_0 > 0$, $\hat{q}_0(0) = 2/(A\lambda_0)$, while for  $ \lambda_0|\mathbf{k}| \gg 1$, $\hat{q}_0 \sim \hat{Q}$.  The similarity between the behaviors of $\hat{q}_0$ and $\hat{Q}$ for large wave numbers in the Fourier domain is naturally related to the fact that $q_0$ and $Q$ have much the same behavior as $r \to 0$. 

It should be emphasized that for $a=0$, the Fourier transform of the reciprocal kernel is a \emph{positive} function given by Eq.~\eqref{46} and this positive character of $\hat{q}$ is expected to persist for sufficiently small $a/\lambda_0 \ll 1$. Indeed, we \emph{suppose} that $a/\lambda_0$ is always so small that $\hat{q}'(\mathbf{k}) > 0$; in fact, this was shown numerically in figure 1 of Ref.~\cite{CM1} for $a/\lambda_0=10^{-3}$. Moreover, figures 1 and 3 of Ref.~\cite{CM1} demonstrate that for $A\lambda_0=10^{-1}$ and $a/\lambda_0=10^{-3}$, $\hat{q}'$ and $\hat{q}''$ have rather similar functional forms: they both start from finite positive values at $|\mathbf{k}| =0$ and monotonically decrease to zero as $|\mathbf{k}| \to \infty$. 

For $a>0$, we find from Eqs.~\eqref{7b},~\eqref{7c} and~\eqref{45} via an integration by parts that 
\begin{equation}\label{47}
\hat{q}'=\hat{q}''-\frac{a}{\lambda_0}\Re \Big [e^Z E_1(Z)\Big]\,,
\end{equation}
and 
\begin{equation}\label{48}
\hat{q}''=\frac{A}{\lambda_0(A^2+|\mathbf{k}|^2)}-\frac{1}{\lambda_0 |\mathbf{k}|}\Im \Big [e^Z E_1(Z)\Big]\,.
\end{equation}
Here 
\begin{equation}\label{49}
Z= a(A+i |\mathbf{k}|)\,,
\end{equation}
$\Re$ and $\Im$ indicate real and imaginary parts, respectively, and $E_1(z)$ is the \emph{exponential integral function} defined for a complex number $z$ with positive real part, $\Re z >0$, as~\cite{A+S}
\begin{equation}\label{50}
E_1(z)= \int_1^{\infty}\frac{e^{-zt}}{t}~dt\,.
\end{equation}

Let us note that $Z$ has a small \emph{positive} real part $aA$, $0<aA\ll 1$, and a \emph{positive} imaginary part $a |\mathbf{k}|$ for $\mathbf{k} \ne 0$. If $\mathbf{k} = 0$, $Z$ is real and $\Im \Big [e^Z E_1(Z)\Big] = 0$; otherwise, one can show---using the lemma connected with Eq. (29) of Ref.~\cite{CM1}---that for complex $Z$, $\Im \Big [e^Z E_1(Z)\Big] < 0$. Moreover, it follows from  $\hat{q}'' > \hat{q}'$  and Eq.~\eqref{47} that $\Re \Big [e^Z E_1(Z)\Big] > 0$. Therefore, $\hat{q}' >0$ whenever
\begin{equation}\label{50A}
0<\frac{a}{\lambda_0}< \frac{\hat{q}''}{\Re \Big [e^Z E_1(Z)\Big]}\,.
\end{equation}
We assume that $a/\lambda_0$ is always so small compared to unity that $\hat{q}' >0$.

 It follows from Eq.~\eqref{50} that in general
\begin{equation}\label{50a}
|e^Z E_1(Z)| \le e^{aA} E_1(aA)\,,
\end{equation}
where equality holds for $\mathbf{k} = 0$.  Moreover,  using the expansion~\cite{A+S} 
\begin{equation}\label{50b}
E_1(z)=-C-\ln z -\sum_{n=1}^{\infty}\frac{(-1)^n z^n}{n~ n!}\,,
\end{equation}
where $C=0.577...$ is Euler's constant, we find that for $0<aA\ll 1$, 
\begin{equation}\label{50c}
E_1(aA) \approx  -C-\ln{(aA)}\,.
\end{equation}
We therefore conclude from Eqs.~\eqref{47} and~\eqref{48} that in general
\begin{equation}\label{50d}
0< \hat{q}'' - \hat{q}' \le \frac{ae^{aA}}{\lambda_0} E_1(aA)\,
\end{equation}
and 
\begin{equation}\label{50e}
0 < \hat{q}'' \le \frac{A}{\lambda_0(A^2+|\mathbf{k}|^2)}+\frac{e^{aA}E_1(aA)}{\lambda_0 |\mathbf{k}|}\,.
\end{equation}

It turns out that for  wavelengths of current observational interest, namely, $\lambda_0|\mathbf{k}| \gtrsim 6\times 10^{4}$, the imaginary part of $Z$ is  much larger than the real part of $Z$; in fact, $\Im Z / \Re Z = A^{-1} |\mathbf{k}| > 6\times 10^{4}$, since $A^{-1} > \lambda_0$ by assumption. It therefore follows from Eqs.~\eqref{50d} and~\eqref{50e} that for gravitational radiation with $\lambda_0|\mathbf{k}| \gtrsim 6\times 10^{4}$, the damping time~\eqref{43a} is always longer than 
\begin{equation}\label{51}
\tau_0= \frac{2\lambda_0|\mathbf{k}|}{Ae^{aA}E_1(aA)}\,.
\end{equation}
For  $A\lambda_0=0.1$ and $a/\lambda_0=0.001$, we can use Eq.~\eqref{50c} with $\ln {10} \approx 2.3$ to conclude that $\tau_0 \gtrsim 5 \times 10^9$ yr, which is about one third of the current estimate for the age of the universe. 

It is interesting that $\tau_0$ is simply proportional to the frequency of gravitational waves. In fact, nonlocality-induced damping could become significant for cosmological gravitational waves with very low frequencies. Inspection of Eq.~\eqref{51} reveals that for realistic values of the parameters $a$ and $A$, $\tau_0$ is longer than the age of the universe for gravitational waves that might be detectable in the foreseeable future with laser interferometers. We recall that for such devices, the waves should have dominant frequency $\gtrsim 1$ Hz for Earth-based and $\gtrsim 10^{-4}$ Hz for space-based antennas. However, detection of gravitational waves with dominant frequency of several nHz may be possible with pulsar timing arrays~\cite{Ri}. Therefore, in connection with waves of low frequency $\nu \sim 10^{-8}$ Hz, we will next examine the limiting cases where $0<a|\mathbf{k}| \ll 1$ and $a|\mathbf{k}|\gg 1$, respectively. 

Consider first the case where $\Im Z = a |\mathbf{k}| \ll 1$; an example is given by $\nu \sim 10^{-8}$ Hz and $a/\lambda_0 = 10^{-8}$. It follows that $|Z| \ll 1$ and to compute $E_1(Z)$ in this case one can use expansion~\eqref{50b}. A detailed examination reveals that in this case
\begin{equation}\label{52}
\hat{q}' \approx \hat{q}'' \approx \frac{\pi}{2\lambda_0|\mathbf{k}|}\,,
\end{equation}
which is much smaller than unity. Eq.~\eqref{52} is reminiscent of the expression~\eqref{45b} for $\hat{Q}$; that is, $a/\lambda_0$ is in this case so small compared to unity as to be essentially negligible.  Hence it follows from Eq.~\eqref{43a} that in this case
\begin{equation}\label{52a}
\tau \approx \frac{\lambda_0|\mathbf{k}|}{A}\,.
\end{equation}
Thus for  $ \lambda_0|\mathbf{k}| \gtrsim 6\times 10^{4}$ and $A\lambda_0=0.1$, we find that $\tau  \gtrsim 2 \times 10^{10}$ yr, which is nearly $1.5$ times the age of the universe. 

Next, we consider the case where $\Im Z = a |\mathbf{k}| \gg 1$. An example is provided by the choice of parameter $a$ in Ref.~\cite{CM1}, namely, $a/\lambda_0=10^{-3}$, so that $\Im Z \gtrsim 60$.  Hence, we need to compute $\hat{q}$ for large $|Z| \gg 1$. It is possible to develop an asymptotic expansion for Eq.~\eqref{50} by using $e^{-zt}~dt=-z^{-1}~d(e^{-zt})$ and repeated integrations by parts. The resulting (divergent) asymptotic expansion for $|z| \gg 1$ is given by~\cite{A+S}
\begin{equation}\label{53}
E_1(z) \sim \frac{e^{-z}}{z}\sum_{n=0}^{\infty}\frac{(-1)^{n} n!}{z^{n}}\,. 
\end{equation}
It thus follows from Eqs.~\eqref{47} and~\eqref{48} that for $|Z|\gg 1$, the dominant terms in $\hat{q}'$ and $\hat{q}''$ are much smaller than unity, and are of the form
\begin{equation}\label{54}
\hat{q}' \sim \frac{4}{\lambda_0 a^3 |\mathbf{k}|^4}\,, \qquad \hat{q}'' \sim \frac{1}{\lambda_0 a |\mathbf{k}|^2}\,.
\end{equation}
Using these asymptotic estimates, we find from Eq.~\eqref{43a} that the corresponding damping times would be
\begin{equation}\label{55}
\tau'  \sim \frac{\lambda_0a^3|\mathbf{k}|^4}{2A}\,, \qquad \tau'' \sim \frac{2\lambda_0a|\mathbf{k}|^2}{A}\,.
\end{equation}
For  $A\lambda_0=0.1$, $a/\lambda_0=0.001$ and $ \lambda_0|\mathbf{k}| \gtrsim 6\times 10^{4}$, one can estimate that $\tau'  \gtrsim 2 \times 10^{15}$ yr and  $\tau'' \gtrsim 2 \times 10^{12}$ yr. Thus in each of the cases considered here, the nonlocality-induced damping of gravitational waves of current observational interest is insignificant as the corresponding damping time is of the order of, or longer than, the age of the universe.

\section{Discussion} 

This paper is devoted to the study of an interesting feature of nonlocal gravity that shows up in the \emph{nonlocal wave equation} for the propagation of free linearized gravitational waves in a background global inertial frame of reference. The waves are \emph{damped} as they propagate in vacuum. The corresponding nonlocal wave equation is reminiscent of a harmonic oscillator with a linear nonlocal damping term. In the case of the simple damped harmonic oscillator, damping or antidamping would depend on the sign of the damping coefficient. A similar situation holds in the nonlocal case, where the nonlocal kernel of the theory must be such that the gravitational waves are all damped and Minkowski spacetime is thus stable. Using certain simplifying assumptions and restricting parameter $a$, $0<a/\lambda_0 \ll 1$, of the kernel to be small enough such that the spatial Fourier transform of the reciprocal Newtonian kernel $\hat{q}$ is positive, we have shown that all the modes do indeed decay. However, for the linearized gravitational waves that may be detectable in the foreseeable future, the amount of damping would be negligible, as the damping time would be of the order of, or longer than, the age of the universe. 

Specifically, the nonlocality-induced damping time $\tau$ is estimated in this paper for radiation of frequency $\nu \gtrsim 10^{-8}$ Hz, which is the frequency range that is the focus of current observational efforts~\cite{Ri}. We recall that $\nu \gg \nu_0$, where, $\nu_0=c/\lambda_0 \approx 10^{-12}$ Hz and $\lambda_0=10$ kpc is the basic galactic length scale associated with nonlocal gravity. It has been shown in a recent work~\cite{CM2} that for $\nu \gtrsim 10^{-8}$ Hz, the nonlocal deviations from standard general relativity are negligibly small, a circumstance that is consistent with our estimate that the corresponding $\tau$ is of the order of, or longer than, the age of the universe. On the other hand, linearized gravitational waves with very low frequencies, $\nu \lesssim \nu_0$, would be highly damped in nonlocal general relativity. 

Finally, it should be emphasized that all of our estimates for the damping time $\tau$ depend, of course, on our assumptions regarding the functional form of the kernel of nonlocal gravity. Ultimately, the kernel of the theory must be determined through comparison with observational data. Therefore, our estimates for $\tau$ may have to be revised as a consequence of the confrontation of nonlocal general relativity with observation.

\begin{acknowledgments}
I am grateful to C.~Chicone and F.~Hehl for valuable discussions. 
\end{acknowledgments}

\end{document}